\documentclass{article}

\usepackage[preprint, nonatbib]{neurips_2020}

\usepackage[utf8]{inputenc} 
\usepackage[T1]{fontenc}    
\usepackage{hyperref}       
\usepackage{url}            
\usepackage{booktabs}       
\usepackage{amsfonts}       
\usepackage{nicefrac}       
\usepackage{microtype}      
\usepackage{amssymb,amsmath,mathrsfs}
\usepackage{comment,multirow,graphicx,bbm,subcaption,caption,multirow}

\usepackage{mathabx,algorithm,algpseudocode}

\newtheorem{theorem}{Theorem}
\newtheorem{definition}{Definition}
\newtheorem{example}{Example}

\title{The Evolutionary Dynamics of Independent Learning Agents in Population Games 
}
\author{
Shuyue Hu$^{\dagger}$, \ Chin-Wing Leung$^{\ddagger}$, \ Ho-fung Leung$^{\ddagger}$, Harold Soh$^{\dagger}$\\
$^{\dagger}$Department of Computer Science, National University of Singapore\\
$^{\ddagger}$Department of Computer Science and Engineering, The Chinese University of Hong Kong\\
$^{\dagger}$\texttt{\{husy,harold\}}@comp.nus.edu.sg\\
$^{\ddagger}$ \texttt{\{cwleung,lhf\}}@cse.cuhk.edu.hk}

\begin{document}

\maketitle

\begin{abstract}
Understanding the evolutionary dynamics of reinforcement learning under multi-agent settings has long remained an  open problem.
While previous works primarily focus on $2$-player games, we consider population games, which model the strategic interactions of a large population comprising small and anonymous agents.
This paper presents a formal relation between stochastic processes and the dynamics of independent learning agents who reason based on the reward signals.
Using a master equation approach, we provide a novel unified framework for characterising population dynamics via a single partial differential equation (Theorem 1).
Through a case study involving Cross learning agents, we illustrate that Theorem 1 allows us to identify qualitatively different evolutionary dynamics, to analyse steady states, and to gain insights into the expected behaviour of a population.
In addition, we present extensive experimental results validating that Theorem 1 holds for a variety of learning methods and population games.
\end{abstract}
\vspace{-0.3cm}
\section{Introduction}
Reinforcement learning (RL) has recently found wide application in real-world multi-agent systems, such as teams of cooperative robots \cite{au2017multirobot,le2017coordinated,long2018towards}, multi-player online games \cite{foerster2017stabilising,justesen2019deep,vinyals2019grandmaster}, distributed sensor networks \cite{colby2013multiagent,gai2018reinforcement,lin2016novel}, and traffic control systems \cite{agogino2012multiagent,casas2017deep,li2016traffic}.
However, in contrast to traditional single-agent settings, a solid theoretical foundation for RL under multi-agent settings remains elusive~\cite{bu2008comprehensive,hernandez2018multiagentsurvey,shoham2007if}.
The fact that multiple autonomous agents interact naturally results in a highly-dynamic environment.
This not only invalidates many convergence guarantees under single-agent settings, but also makes the learning dynamics much more complex and unpredictable~\cite{tuyls2007evolutionary}.
The latter, usually referred to as non-stationarity, is a key challenge in multi-agent reinforcement learning (MARL)~\cite{hernandez2017survey}.
In light of the above issues, 
it is of both theoretical and practical interest to develop a thorough understanding of the evolutionary learning dynamics in multi-agent systems.
Understanding the dynamics may facilitate important tasks, such as selecting a specific algorithm for a given problem~\cite{bloembergen2015evolutionary}, shedding light on the design of a new algorithm~\cite{balduzzi2018mechanics,kaisers2010frequency}, and parameter tuning~\cite{panait2008theoretical,tuyls2003selection}. More broadly, the gained insights can engender a more \emph{trustworthy} AI ecosystem.  

In general, prior work on MARL dynamics has primarily focused on $2$-player games~(e.g., \cite{bailey2019multi,bloembergen2015evolutionary,gatti2013efficient,hu2019modelling,kaisers2010frequency,klos2010evolutionary,tuyls2005evolutionary}). 
However, many social, economic and technological scenarios involve a large number of agents, in which each agent's payoffs largely rely on the aggregated effect of the other agents. For example, consider drivers commuting over a highway network; the congestion each driver experiences depends not only on the route he selects, but also on the number of drivers along the same route.
As another example, the choice of a social networking platform often depends on the choices of one's friends.
Population games \cite{sandholm2010population} provide a \emph{unified} framework to model and to analyse the above situations comprising a great number of agents.
However, despite wide application, the learning dynamics in population games has rarely been investigated.

This paper aims to formulate the evolutionary dynamics of RL in population games.
The large number of agents in the games calls for an efficient formulation---deriving a dynamical system for each agent is impractical for a large population.
Here, we take a probabilistic viewpoint: we model, at given time, the critical parameter(s) that characterise an agent's learning by a random variable, such that the probability density of a parameter is asymptotically its frequency in the population.
Over time, the collection of random variables forms a stochastic process.
We find that understanding this relation between MARL and stochastic processes is crucial, as it opens new ways of analysing the evolutionary learning dynamics.
In particular, by using a master equation~\cite{scott2013applied}, we show how the time evolution of the probability distribution of the random variable describes the learning dynamics in population games.
As a result, the learning dynamics of a large number of agents can be efficiently formulated by a \emph{single} partial differential equation (PDE).

In this paper, we focus our attention to the class of \emph{independent learning} algorithms in multi-agent systems. By assuming agents reason based on the reward signals, this line of research effectively reduces the multi-agent learning problem to a single-agent one~\cite{bloembergen2015evolutionary,hernandez2017survey,2012matignonindependent}.
The benefits are twofold: (i) the vast number of single-agent learning methods can be directly applied to multi-agent settings, which has been shown to achieve good performance in many multi-agent systems \cite{abed2016comparison,bloembergen2015evolutionary,bu2008comprehensive,hernandez2018multiagentsurvey,2012matignonindependent}.
More importantly, (ii) scalability in the number of other agents in a system is no longer an issue.
This makes it particularly well-suited to population games that involve a great number of agents.

We present an abstract representation of independent learning agents, and establish its relation with stochastic processes. This leads to a unified framework (Theorem 1) for formulating the evolutionary dynamics in population games for a variety of independent learning methods. 
Instantiating Theorem 1 for each learning algorithm leads to a single PDE that characterises the dynamics of a homogeneous agent population using that algorithm.
We illustrate through a case study (on Cross learning \cite{cross1973stochastic}) how Theorem 1 helps us identify the qualitatively different evolutionary dynamics, analyse the steady states, and understand the trend of the expected behaviours of a population.
In addition, experiments with two additional well-known learning methods ($Q$-learning \cite{watkins1992q} and infinitesimal gradient ascent \cite{singh2000nash}) demonstrate that our approach well-describes the learning dynamics observed in agent-based simulations for various population games.

\section{Background and Related Work}
In this section, we present a brief overview of population games and independent learning.
\subsection{Population Games}
Recall the examples of highway congestion and social network choice in the introduction. While these two situations differ, they have the same basic properties in common: (i) the number of agents is large, (ii) each agent is small, such that any particular one agent’s behaviour has little or negligible effect on other individual agents, and (iii) each agent is anonymous, in that exchanging the labels of agents will not create any difference. 
Population games \cite{sandholm2010population} provide a unified framework to model situations simultaneously exhibiting the above three properties. 
In population games, an agent's rewards depend only on its own behaviour and the aggregate effect of many other agents' behaviours, and is independent of its identity (or label). 
Formally, we define a population game as follows.

\begin{definition}[Population Games]
\label{def:game}
A population  game $G$ is a tuple $\langle \mathcal{N}, \mathcal{A}, \mathcal{O}, R \rangle$. $\mathcal{N}=\{1, \ldots, n\}$ is a large set of $n$ agents.
$\mathcal{A}=\{a_1, \ldots ,a_k\}$ is a set of $k$ actions available to each agent.
Let $\mathbf{o}=[o_1, \ldots,\ o_k]^\top$ be a population profile (or outcome), where $o_i$ denotes the proportion of agents choosing action $a_i$ in the population.
$\mathcal{O}=\{\mathbf{o} \in \mathbb{R}^k_+: \sum_{i=1}^{k} o_i =1\}$ is a set of population profiles. 
$R(a, \mathbf{o}):\mathcal{A} \times \mathcal{O} \to \mathbb{R}$ is a reward function that determines the reward of an agent by the action $a\in \mathcal{A}$ it uses and the population profile $\mathbf{o}\in \mathcal{O}$. 
\end{definition}

We consider that agents learn their strategies through the repeated plays of a population game.
For each time step $t$, all of the agents take an action simultaneously, receive immediate rewards, and update their strategies.
At the next time step $t+1$, agents start over for another play of the game. 
Therefore, there is no explicit state transition in each play of the population game, just as in other normal form games.
During repeated plays, what an agent learns is which action to choose in each play of the game.

We note that as the number $n$ of agents tends to infinity, one may view population games to be mean field games \cite{huang2006large,lasry2007mean}. 
We deliberately use the term population games, since mean field games are typically associated with optimal agent control; agents do not learn in mean field games other than in a few notable exceptions~\cite{guo2019learning,mguni2018decentralised,subramanian2019reinforcement}. 
However, these works aim for the final convergence to Nash equilibriua in the time limit $t\to \infty$, while we emphasise the \emph{dynamic process} through which agents adjust their behaviours in response to the strategic agent population.

\subsection{Independent Learning and Critical Parameters}
Independent learning agents reason based on only the reward signals, and do not perceive or utilise information about other agents.\footnote{In many practical applications, it is not reasonable to assume the observability of other agents; most agents rely on
sensory information and action recognition that is often far from trivial~\cite{2012matignonindependent}. }
The interaction of an independent learner with an environment is typically defined as a Markov decision process (MDP). 
A MDP consists of a set $\mathcal{S}$ of states, a set $\mathcal{A}$ of actions, an immediate reward function $R(s,a)$ giving the reward of using action $a$ under state $s$, and a transition function $T(s'|s,a)$ that determines the probability of a transition from a state $s$ to another state $s'$ after using action $a$. There is no explicit state transition in a population game; from an agent's point of view, it stays in the same state $s$ during the repeated plays.
As such, the interaction with the environment is a single-state (or stateless) MDP.
Other agents affects an independent learner only through the immediate rewards the learner receives.

It is interesting to note that most RL algorithms repeatedly apply an update rule to a set of \emph{critical parameters}.
Consider 
$Q$-learning \cite{watkins1992q}, one of the most well known value-based independent learning methods.
In a stateless MDP, a $Q$-learning agent maintains a vector of $Q$-values $\mathbf{Q}(t)=[Q_1(t), \ldots, Q_k(t)]^\top$, each of which $Q_j(t)$ estimates the expected reward of using a particular action $a_j \in \mathcal{A}$.\footnote{Note that the dependency of $Q$-values on state $s$ is dropped, since there is only one state $s$.}
At every time step $t$, the agent updates the $Q$-values based on the immediate reward, and then updates its policy (i.e., the probability of taking each action) based on a certain exploration strategy. 
For a $Q$-learning agent, the time evolution of both the $Q$-values and the policy can well characterise the agent's learning dynamics \cite{hu2019modelling,Gomes2009dynamic,tuyls2003selection,wunder2010classes}.  
Therefore, there are two types of critical parameters in $Q$-learning: the $Q$-values and the policy.

Tuyls et al. \cite{tuyls2003selection} focus on the policy of a $Q$-learning agent, and derive two differential equations each for one agent, which describe the time evolution of the policies in two-player games.
More recently, Hu et al. \cite{hu2019modelling} focus on the $Q$-values, and show that a differential equation for an agent's $Q$-values can also be derived.
In general, 
a significant number of works 
\cite{bloembergen2010lenient,borgers1997learning,galstyan2013continuous,gatti2013efficient,kaisers2012common,klos2010evolutionary,panait2008theoretical,Gomes2009dynamic,wunder2010classes} have shown that the learning dynamics of an individual agent can be formulated by a differential equation of the critical parameters.
However, the choice of critical parameters may vary in different learning methods and out of interest.
We refer readers to \cite{bloembergen2015evolutionary} for a comprehensive survey.

Our work builds upon this line of research. 
In particular, our main result is that 
under certain conditions, the population dynamics of independent learners can be described by a PDE, in which an ODE characterising the change in the critical parameters of a single agent is incorporated.

\section{Modelling the Population Dynamics of Independent Learners}
\label{sec:model}
Consider a large set $\mathcal{N}$ of $n$ independent learners that play a population game repeatedly. 
For any agent $i\in \mathcal{N}$, let 
$\mathbf{x}^{(i)}(t) = [x^{(i)}_1(t),  \ldots,  x^{(i)}_d(t) ]^\top \in \mathbb{R}^d$  be a vector of $d$ independent critical parameters at time $t$, to which the agent repeatedly applies an update rule.
The dimension $d$ indicates the degree of freedom.

To model the population dynamics, one may consider a straightforward system of $n$ differential equations, each of which characterises the change in the critical parameters of one agent.
This approach may be feasible for a small number of agents.
However, the differential equations are coupled through the immediate rewards because agents affect one another through the rewards.
As the number of agents grows, the corresponding system of $n$ coupled differential equations becomes intractable.

Alternatively, one can take a probabilistic point of view.
At time $t$, consider an arbitrary agent that is randomly drawn from the agent population.
The critical parameters of this agent can then be modelled by a $d$-dimensional random variable $\mathbf{X}(t)= [X_1(t), \ldots, X_d(t)]^\top \in \mathbb{R}^d$ with a probability density function (PDF) $p(\mathbf{x},t)$ where $\mathbf{x}$ is a realisation of the random variable $\mathbf{X}(t)$.
Let $M_n(\mathbf{x},t)$ be the empirical cumulative distribution function (CDF) of the critical parameters in the population, i.e.,
\begin{equation}
 M_n(\mathbf{x},t)  \triangleq \frac{1}{n} \sum_{i\in \mathcal{N}} \mathbbm{1} (\mathbf{x}^{(i)}(t)\leq \mathbf{x}),
\end{equation}
where the indicator function $\mathbbm{1}(\mathbf{x}^{(i)}(t)\leq\mathbf{x})$ equals $1$ if $x^{(i)}_j(t)\leq x_j$ for $j\in \{1,\ldots, d\}$, or equals $0$ otherwise. 
We define the CDF $P(\mathbf{x},t)$ of the random variable $\mathbf{X}(t)$ as the asymptotic distribution of the empirical CDF $M_n(\mathbf{x},t)$, such that
\begin{equation}
 M_n(\mathbf{x},t) \overset{\mathcal{D}}{\to} P(\mathbf{x},t) \quad \text{and} \quad p(\mathbf{x},t)=\frac{d P(\mathbf{x},t)}{d\mathbf{x}}.
\end{equation}
Therefore, as the number of agents goes to infinity, at time $t$, the probability density $p(\mathbf{x},t)$ can be intuitively interpreted as the proportion of agents having the particular vector $\mathbf{x}$ of critical parameters in the population.

Over time, the collection of random variables $\mathbf{X}(t)$ forms a stochastic process $\{\mathbf{X}(t); t\in \mathbb{R}^+\}$.
In the following, we present and explain our main result in Section 3.1.
We then show in Section 3.2 how we use a master equation approach---a technique of stochastic process \cite{scott2013applied}---to derive the main result.
For ease of analysis, our working assumption is $n \to \infty$.\footnote{Interestingly, experiments show that the learning dynamics of a population of only $1,000$ agents can be well described by our approach.}

\subsection{Main Result: a Partial Differential Equation Describing the Population Dynamics}
\label{sec:result}
We consider that at time $t$, each independent learning agent $i \in \mathcal{N}$ takes an action, receives an immediate reward $r^{(i)}(t)$, and may have a set $\mathcal{I}(t)$ of additional information that is homogeneous for every agent in the population.
The information set $\mathcal{I}(t)$ is typically an empty set indicating that an agent has no information other than the immediate reward it receives.
However, it is possible for independent learning agents to have access to some additional information, such as the possible rewards of using other actions (e.g., in gradient ascent \cite{kaisers2012common}), and the optimal reward in hindsight (e.g., in regret minimisation \cite{blum2007external}), which do not explicitly reveal the existence of other agents.
Suppose every agent updates its critical parameters by making use of the immediate reward and the information set.
Our main result is summarised in the following theorem.
\begin{theorem}
Consider a set $\mathcal{N}$ of agents, each using the same independent learning method.
If there exists a vector-valued function $f_{\theta}=[f_{\theta,1}, \ldots,  f_{\theta,d}]^\top$ and a function $g$,  such that for every agent $i \in\mathcal{N}$,
\begin{align}
   \frac{d\mathbf{x}^{(i)}(t)}{dt}&=f_{\theta}(\mathbf{x}^{(i)}(t),r^{(i)}(t),\mathcal{I}(t),t), 
   \label{eq:differential} \\
     r^{(i)}(t) &= g(\mathbf{x}^{(i)}(t),t),
   \label{eq:g}
\end{align}
and $f_{\theta,j}$ is differentiable at any  $x_j^{(i)}(t)$  for all $j\in \{1, \ldots, d\}$, then the population dynamics is described by the time evolution of the probability distribution $p(\mathbf{x},t)$ given as follows: 
\begin{equation}
\begin{aligned}
     \frac{\partial p(\mathbf{x},t)}{\partial t} &= - \sum_{j=1}^{d} \frac{\partial } {\partial {x_j}} [p(\mathbf{x},t) f_{\theta,j}(\mathbf{x},g(\mathbf{x},t),\mathcal{I}(t),t)].
     \end{aligned}
    \label{eq:mainresult}
\end{equation}
\end{theorem}
Note that the density $p(\mathbf{x},t)$ is intuitively the proportion of agents having the vector $\mathbf{x}$ of critical parameters in the population.
Therefore, Equation \ref{eq:mainresult} effectively shows how the proportion of agents having each possible vector of critical parameters evolves as time goes forward.

Equation \ref{eq:differential} characterises the time evolution of the critical parameters of a single agent $i$, in which the function $f_{\theta}$ represents the update rule that is repeatedly applied to the critical parameters.
The subscript $\theta$ represents the set of non-critical parameters used in the update rule, e.g., the learning rate, discount factor, and Boltzmann exploration temperature.
In contrast to the critical parameters, they are given at the outset, unchanged over time, and are the same for every agent in the population.

The critical parameters should exhibit Markov property by the function $f_{\theta}$.
 Equation \ref{eq:mainresult} requires the partial derivative of $f_{\theta,j}$ with respect to $x_j$ to exist for each dimension $j\in \{1,\ldots, d\}$.
The function $g$ stipulates that at any time $t$, there is exactly one reward for a particular vector of critical parameters.

The most noteworthy aspect of this theorem is that it reduces the problem of formulating the learning dynamics of an agent population to solving a single PDE (Equation \ref{eq:mainresult}).
This is much more tractable compared with a system of $n$ coupled differential equations.
Moreover, Theorem 1 does not explicitly rely on a particular learning algorithm; Equation \ref{eq:differential}, the differential equation describing the learning dynamics of a single agent, can be instantiated for various learning methods.
Therefore, building upon the line of research that derives the differential equation for individual agents using a specific learning method~\cite{bloembergen2010lenient,bloembergen2015evolutionary,borgers1997learning,hu2019modelling,kaisers2012common,panait2008theoretical,tuyls2003selection}, 
we will be able to show, for the first time, the population dynamics of various methods in a unified framework.

\subsection{Derivation: a Master Equation Approach for Homogeneous Agent Populations}
In the following, we give a brief derivation of Theorem 1; please see the appendix for more details. As agents update their critical parameters during learning, the PDF $p(\mathbf{x},t)$ evolves in time.
Let us consider the change in $p(\mathbf{x},t)$ within an infinitesimal time interval $(t,t+ dt)$. Two cases account for the change during the time interval:
(i) agents with parameters $\mathbf{x}'$ change their parameters to $\mathbf{x}$, and 
(ii) agents with parameters $\mathbf{x}$ change their parameters to $\mathbf{x}'$.
Other cases that involve multiple steps, such as changing from $\mathbf{x}$ to $\mathbf{x}'$, and then back to $\mathbf{x}$ again, need not to be considered, as $dt\to 0$.
The above contributions to the change in $p(\mathbf{x},t)$ can be formulated by a master equation~\cite{siegert1949approach}:
\begin{equation}
\label{eq:master}
    \frac{\partial p(\mathbf{x},t)}{\partial t} = \int \left[T(\mathbf{x}|\mathbf{x}',t)p(\mathbf{x}',t) - T(\mathbf{x}'|\mathbf{x},t)p(\mathbf{x},t)]\right. d\mathbf{x}',
\end{equation}
where $T(\mathbf{x}|\mathbf{x}',t)$ is the probability of a transition from $\mathbf{x}'$ to $\mathbf{x}$ at time $t$ (and likewise for $T(\mathbf{x}'|\mathbf{x},t)$). A master equation requires the random variable to exhibit Markov property \cite{siegert1949approach}, so that the transition probabilities can be written in the forms $T(\mathbf{x}|\mathbf{x}',t)$ and $T(\mathbf{x}'|\mathbf{x},t)$. Equation \ref{eq:differential} in Theorem 1 provides this guarantee. 
Since independent learning methods are typically defined by MDPs, the critical parameters generally satisfy this property.

Nevertheless, one may ask how we can ensure the time evolution of the $p(\mathbf{x},t)$ is asymptotically that of the empirical distribution.
Consider the end of the time interval, $t+dt$.
Suppose for any agent $i\in \mathcal{N}$, the function $\mathcal{T}^{(i)}(\mathbf{x}|\mathbf{x}',t)$ determines the probability of its transition from the parameters $\mathbf{x}'$ to $\mathbf{x}$ during the time interval.
We define $\Delta M_n(\mathbf{x},t)$ as the proportion of agents whose parameters are within the interval $(\mathbf{x}-d \mathbf{x}, \mathbf{x}]$ at time $t$. That is, $\Delta M_n(\mathbf{x},t) \triangleq M_n(\mathbf{x},t) - M_n(\mathbf{x}-d \mathbf{x},t)$ where $M_n$ is the empirical CDF of $\mathbf{X}(t)$.
Since the critical parameters satisfy the Markov property, 
by the Chapman–Kolmogorov equation \cite{chapman1928brownian,kolmogoroff1931analytischen},
\begin{equation}
\begin{aligned}
 \Delta M_n(\mathbf{x},t+dt)  
=    \frac{1}{n} \sum_{i\in \mathcal{N}} \int \mathbbm{1} (\mathbf{x}^{(i)}(t) \in (\mathbf{x}'-d\mathbf{x}, \mathbf{x}' ]) \mathcal{T}^{(i)}(\mathbf{x}|\mathbf{x}',t) d\mathbf{x}' 
= \int \Delta M_n(\mathbf{x}',t) \mathcal{T}^{(i)}(\mathbf{x}|\mathbf{x}',t)  d\mathbf{x}'.
 \end{aligned}
\end{equation}
Similarly, $p(\mathbf{x},t+dt)= \int p(\mathbf{x}',t) T(\mathbf{x}|\mathbf{x}',t)  d\mathbf{x}'$. To ensure $M_n(\mathbf{x},t+dt) \overset{\mathcal{D}}{\to} P(\mathbf{x},t+dt)$, the following equation should hold:
\begin{equation}
T(\mathbf{x}|\mathbf{x}',t) = \mathcal{T}^{(i)}(\mathbf{x}|\mathbf{x}',t), \qquad \forall i\in \mathcal{N}.
\end{equation}
At any time $t$, the transition probability $\mathcal{T}^{(i)}(\mathbf{x}|\mathbf{x}',t)$ should be the same for every agent in the population. 
In essence, this requires that agents should be homogeneous, in that the transition probability is always uniquely determined by the critical parameters, and is independent of the identity of agents.
This requirement accounts for the condition that every agent in the population use the same learning method, in which the non-critical parameters are the same and the additional information is homogeneous.
Moreover, there should exist a function (function $g$ in Theorem 1), such that the immediate reward an agent receives is only determined by its critical parameters. In other words, agents having the same critical parameters receive exactly the same immediate reward.

A master equation in its generic form provides little information without instantiating the transition probability.
Let $\mathbf{x}'$ be $\mathbf{x}+\Delta \mathbf{x}$.
We denote $T(\mathbf{x}'|\mathbf{x},t)$ by $ \mathscr{T}(\mathbf{x},\Delta \mathbf{x},t)$ such that 
the latter is the transition probability from $\mathbf{x}$ to $\mathbf{x}+\Delta \mathbf{x}$ at time $t$.
Similarly, denote $T(\mathbf{x}|\mathbf{x}',t)$ by
$\mathscr{T} (\mathbf{x}+\Delta \mathbf{x},-\Delta\mathbf{x},t) $. 
Then, Equation \ref{eq:master} can be rewritten as
\begin{equation}
\begin{aligned}
    \frac{\partial p(\mathbf{x},t)}{\partial t} &=\int \left [\mathscr{T} (\mathbf{x}+\Delta \mathbf{x},-\Delta\mathbf{x},t) p(\mathbf{x}+\Delta \mathbf{x},t) -  \mathscr{T} (\mathbf{x},\Delta\mathbf{x},t) p(\mathbf{x},t)] \right. d(\mathbf{x}+\Delta \mathbf{x}),\\
 \end{aligned} 
 \label{eq:master1}
\end{equation}
where $d(\mathbf{x}+\Delta \mathbf{x})$ can be replaced with $d\Delta \mathbf{x}$. Using a Taylor expansion on the first term on the RHS,
\begin{equation}
\begin{aligned}
   \frac{\partial p(\mathbf{x},t)}{\partial t} &=  \int \left[\mathscr{T} (\mathbf{x},-\Delta\mathbf{x}, t) p(\mathbf{x},t) 
    +\Delta \mathbf{x} \cdot \nabla_{\mathbf{x}} \left[\mathscr{T}(\mathbf{x},-\Delta \mathbf{x},t) p(\mathbf{x},t)\right ]
     - \mathscr{T} (\mathbf{x},\Delta\mathbf{x},t) p(\mathbf{x},t)]\right.
     d\Delta \mathbf{x}.
 \end{aligned}
 \label{eq:master2}
\end{equation}
Now, we instantiate the transition probabilities $ \mathscr{T}(\mathbf{x}, \Delta \mathbf{x},t)$ and $ \mathscr{T}(\mathbf{x}, -\Delta \mathbf{x},t)$.
Recall that for each agent, the time evolution of its critical parameters is given by Equation \ref{eq:differential}. 
At given time $t$, the probability of a transition is deterministic and can be defined with a delta function, 
\begin{equation}
\begin{aligned}
    \mathscr{T}(\mathbf{x}, \Delta \mathbf{x},t)  \triangleq \delta(f_{\theta}(\mathbf{x},g(\mathbf{x},t),\mathcal{I}(t),t) - \Delta \mathbf{x}).
\end{aligned}
\end{equation}
The transition probability $\mathscr{T}(\mathbf{x}, \Delta \mathbf{x},t) \to +\infty$ if the change in $\mathbf{x}$ at time $t$ (given by $f_{\theta}(\mathbf{x},g(\mathbf{x},t),\mathcal{I}(t),t)$) is exactly $\Delta \mathbf{x}$.
Otherwise, the probability is $0$.
Similarly, $\mathscr{T}(\mathbf{x}, -\Delta \mathbf{x},t)
    \triangleq \delta(f_{\theta}(\mathbf{x},g(\mathbf{x},t),\mathcal{I}(t),t) + \Delta \mathbf{x})$.
Note that $\int \mathscr{T} (\mathbf{x},\Delta\mathbf{x},t) d\Delta \mathbf{x} = \int \mathscr{T} (\mathbf{x},-\Delta\mathbf{x},t) d\Delta \mathbf{x} =1$.
The first and the third term in the integrand of Equation \ref{eq:master2} can be cancelled out, giving
\begin{equation}
\begin{aligned}
   \frac{\partial p(\mathbf{x},t)}{\partial t} & =
   \int  \Delta{\mathbf{x}} \cdot \nabla_{\mathbf{x}}[\delta(f_{\theta}(\mathbf{x}, g(\mathbf{x},t) ,\mathcal{I}(t),t) + \Delta \mathbf{x}) p(\mathbf{x},t)]  d\Delta \mathbf{x}. \\
 \end{aligned}
 \label{eq:master3}
\end{equation}
As presented in the appendix, Equation \ref{eq:master3} will yield
\begin{equation}
\begin{aligned}
     \frac{\partial p(\mathbf{x},t)}{\partial t} &= - \sum_{j=1}^{d} \frac{\partial } {\partial {x_j}} \left[p(\mathbf{x},t) f_{\theta,j}(\mathbf{x}, g(\mathbf{x} ,t),\mathcal{I}(t),t)\right], \\
     \end{aligned} 
     \nonumber
\end{equation}
which is  Equation \ref{eq:mainresult} presented in Theorem 1. We note that Equation \ref{eq:mainresult} can be viewed as a special case of the Fokker-Planck-Kolmogorov equation~\cite{fokker1914mittlere,kolmogoroff1936anfangsgrunde,planck1917satz} with drift function $f_{\theta}=(\mathbf{x}, g(\mathbf{x} ,t),\mathcal{I}(t),t)$ and zero diffusion.

\section{A Case Study: Cross Learning Agents}
In this section, we illustrate the application of Theorem 1 by focusing on Cross learning \cite{cross1973stochastic}, which is arguably the origin of reinforcement learning \cite{borgers1997learning}, and 
a typical prototype for studying learning dynamics \cite{bloembergen2015evolutionary}.
Let us consider an infinitely large population of Cross learning agents, each of which has two available actions $\mathcal{A}=\{a_1, a_2\}$.
A Cross learning agent maintains a vector of mixed-strategy policy $\boldsymbol{\pi}(t)= [\pi_1(t), \pi_2(t)]^\top$, in which each element is the probability of using an action. 
At time $t$, if an agent takes action $a_j \in \mathcal{A}$ and receives an immediate reward $r(t)$, then it will update its policy $\boldsymbol{\pi}(t)$, as follows:
\begin{equation} 
\pi_i(t+1)= \pi_i(t) +
    \begin{cases}
    r(t) - \pi_i(t) \times r(t) & \text{if } a_i=a_j\\
  - \pi_i(t)  \times r(t)  & \text{else} \\
    \end{cases}, \qquad \forall a_i \in \mathcal{A}.
    \label{eq:cross}
\end{equation}
Without loss of generality, we consider the probability $\pi_{1}(t)$ of taking the first action $a_1$ to be the critical parameter; the degree $d$ of freedom is $1$ here, since $\pi_1(t)+\pi_2(t)=1$.
For a single agent, the time evolution of the probability of taking an action (which is $a_1$ here) can be described by the following differential equation~\cite{borgers1997learning}:
\begin{equation}
\frac{d\pi_1(t)}{dt}
=  \pi_1(t) \left[r(t)-\sum_{j: a_j\in \mathcal{A}}\pi_j(t) r_j(t)\right] = \pi_1 (1-\pi_1) \left[r_1(t) -r_2(t)\right],
\label{eq:crossDif}
\end{equation}
where $r_j(t)$ is the reward at time $t$ if action $a_j$ is used.
By Theorem 1, the time evolution of the learning dynamics in this population is then given by 
\begin{equation}
\frac{\partial p(\pi_1,t)}{\partial t} 
= - \frac{\partial } {\partial {\pi_1}} \left[p(\pi_1,t) \frac{d\pi_1(t)}{dt}\right].
\label{eq:crossfpk}
\end{equation}

\begin{figure}
\begin{subfigure}{0.24\textwidth}
\centering
  \includegraphics[width=0.95\textwidth]{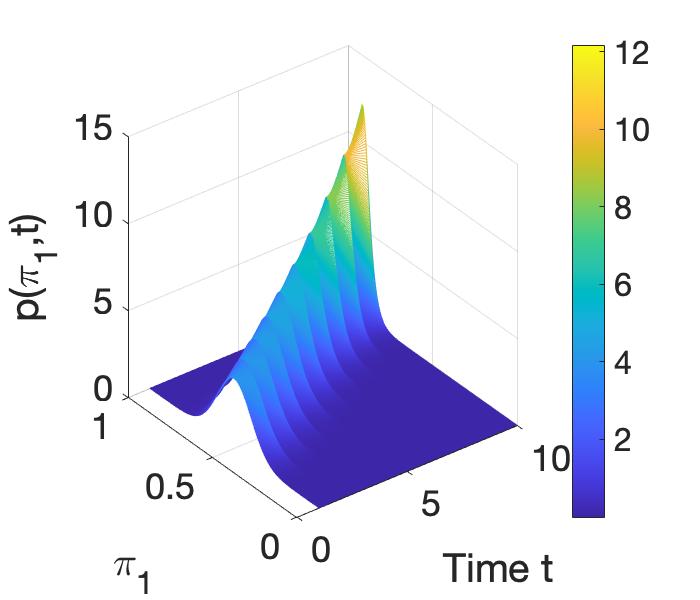}
  \small
  \caption{\small $\pi_1\sim \textsc{N}(0.5,0.1^2)$}
  \label{fig:1a}
  \end{subfigure}
  \begin{subfigure}{0.24\textwidth}
  \centering
  \includegraphics[width=0.95\textwidth]{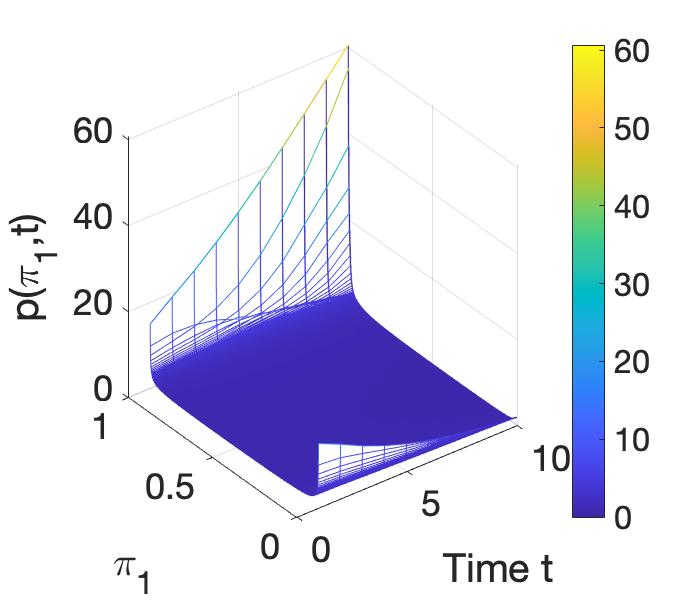}
  \small
  \caption{\small  $\pi_1 \sim \text{Beta}(0.4,0.4)$}
  \label{fig:1b}
  \end{subfigure}
  \begin{subfigure}{0.24\textwidth}
  \centering
  \includegraphics[width=0.95\textwidth]{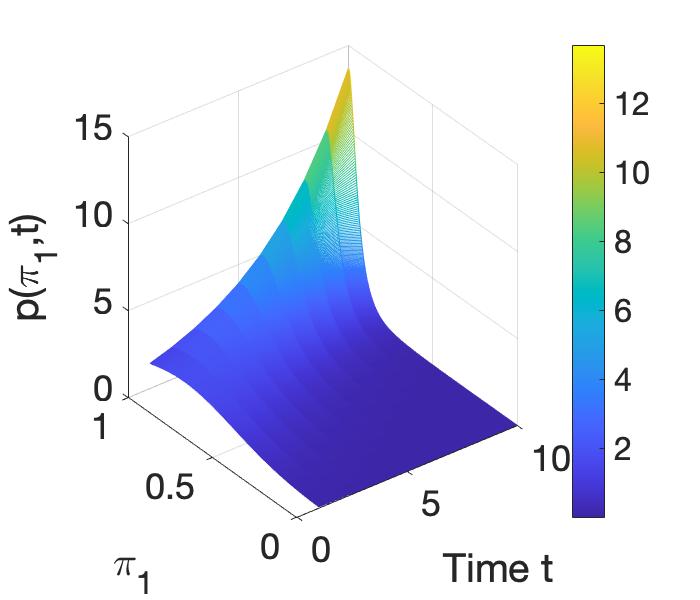}
  \small
  \caption{\small $\pi_1 \sim \textsc{N}(0.8,0.3^2)$ }
  \label{fig:1c}
  \end{subfigure}
    \begin{subfigure}{0.24\textwidth}
  \centering
  \includegraphics[width=0.95\textwidth]{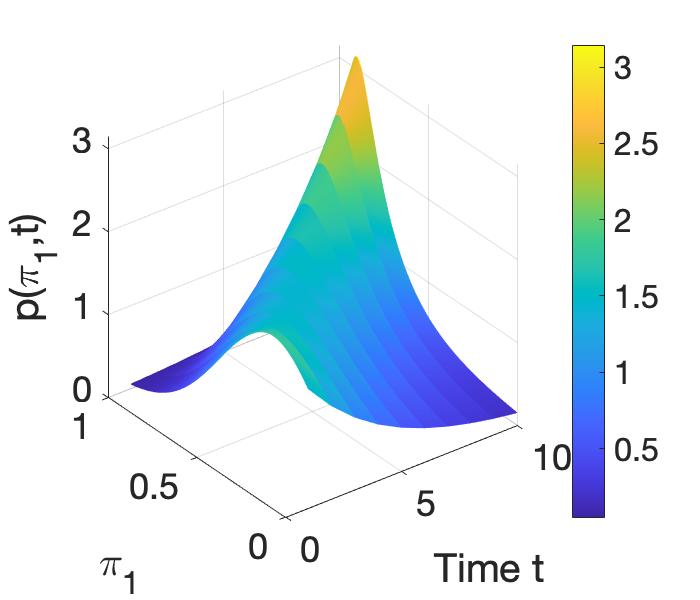}
  \small
  \caption{\small $\pi_1 \sim \textsc{N}(0.2,0.3^2)$}
  \label{fig:1d}
  \end{subfigure}
    \caption{The evolution of the $p(\pi_1,t)$ in public goods games with different initial distributions. }
      \vspace{-0.3cm}
\end{figure}

\paragraph{Identifying qualitatively different evolutionary dynamics.}
The main usefulness of Equation \ref{eq:crossfpk} is that it allows us to gain insight into the evolutionary learning dynamics in population games, without computationally-expensive agent-based simulations.
For ease of demonstration, consider the well-known public goods (or $n$-player prisoner's dilemma) game. 
In this game, each agent has two actions: defect (denoted by $a_1$) or cooperate (denoted by $a_2)$. Let $\psi$ be the current proportion of cooperating agents at time $t$.  
An agent will receive a reward $r_1(t)=1.5\psi$ if it defects, or a reward $r_2(t)=1.5\psi-0.5$ if it cooperates.
Suppose at time $t=0$, the probability $\pi_1$ of taking action $a_1$ (defection) for each agent is distributed according to the truncated normal distribution $\mathsf{N}(0.5,0.1^2)$.

By solving Equation \ref{eq:crossfpk} with such initial distribution,\footnote{In this paper, we use finite difference to solve the PDEs.}
one can obtain how $p(\pi_1,t)$ evolves over time (shown in Figure 1(a)).
It is unsurprising that as time proceeds, the population converges to defect with high probability, because defection always results in a higher reward than cooperation.
However, as shown in Figures 1(a)-1(d), with different initial distributions, the time evolution of $p(\pi_1,t)$ can be qualitatively different.
Such information is usually elusive in agent-based simulations, given the randomness arising from each simulation run.
In contrast, our approach immediately discloses this information, simply  
by solving Equation \ref{eq:crossfpk}.

\paragraph{Understanding the trend of expected behaviours.}
Another application of Equation \ref{eq:crossfpk} is to investigate the expected behaviours of agents in the population.
Let $\mathbb{E}[\pi_1(t)]$ be the expected probability of taking action $a_1$ in the population  at time $t$. It can be shown that
\begin{equation}
    \begin{aligned}
    \frac{d\mathbb{E}[\pi_1(t)]}{dt} &= \frac{d}{dt} \int_{0}^{1} \pi_1 p(\pi_1,t) d\pi_1 
= [r_2(t)-r_1(t)] \int_{0}^{1} \pi_1 (\pi_1-1) p(\pi_1,t) d\pi_1.
    \end{aligned}
\end{equation}
The integral on the RHS is smaller than $0$. Hence, at time $t$, the value of $\frac{d\mathbb{E}[\pi_1(t)]}{dt}$ will be positive if $r_2(t)<r_1(t)$, or be negative if $r_2(t)>r_1(t)$.
This indicates that 
the expected probability $\mathbb{E}[\pi_1(t)]$ of taking action $a_1$ changes in the direction of a higher immediate reward.
Moreover, if the distribution $p(\pi_1,t)$ is more concentrated around $\pi_1=0.5$, the magnitude of the integral on the RHS will increase. That is, with a larger proportion of agents not showing an explicit tendency to choose a particular action, the expected change in their probability of taking each action becomes more intense.

\paragraph{Steady state analysis.} 
Though it is not our intention to analyse the convergence to Nash equilibria, Equation \ref{eq:crossfpk} also allows us to study steady states.
Consider $\frac{\partial p(\pi_1,t)}{\partial t}=0$, such that there is no change in $p(\pi_1,t)$ of each realisation $\pi_1$, and thus the distribution is completely steady.
Then, 
\begin{equation}
 [r_1(t) -r_2(t)] \frac{\partial p(\pi_1,t) }{\partial \pi_1} 
=  -[r_1(t) -r_2(t)]  \frac{1-2\pi_1}{(1-\pi_1)\pi_1} p(\pi_1,t).
\label{eq:steady}
\end{equation}
Solving this equation with the property $\int_0^1 p(\pi_1,t) d\pi_1 =1$, we have $p^\ast(\pi_1,t)=\frac{c}{\pi_1(1-\pi_1)}$ with $c\to0^+$. In this steady distribution, we have 
 $p^\ast(\pi_1,t)\to +\infty$ when $\pi_1 \in\{0,1\}$, and $p^\ast(\pi_1,t)\to0$ when $\pi_1 \in(0,1)$.
This indicates that half of the agents in the population have $\pi_1=0$, while the other half have $\pi_1=1$, regardless of their immediate rewards. 
This is confirmed by Equation \ref{eq:cross}---a Cross learning agent's policy remains unchanged once $\pi_1$ reaches $0$ or $1$.
Therefore, in fact, there are countless possible steady states, in which each agent takes either $\pi_1=0$ or $\pi_1=1$, yet which particular value an agent takes does not matter.
Once the population evolves into one of these states or takes one of these states as the initial distribution, the population will remain in that state forever.

\section{Validation on Various Learning Methods and Population Games}
In this section, we focus on validating our approach across different learners and population games. We selected three different learning methods, i.e., Cross learning \cite{cross1973stochastic}, $Q$-learning~\cite{watkins1992q} with Boltzmann exploration, and Infinitesimal Gradient Ascent (IGA) \cite{singh2000nash}.
In contrast to Cross learning and $Q$-learning, IGA assumes that agents know the expected reward function ($\mathcal{I}(t)$ is non-empty).\footnote{IGA was originally designed for two-player-two-action games. We extend IGA by assuming that the expected reward of taking each action is known to agents. Details are provided in the appendix.}
For comparison, we considered five configurations for three typical population games: the public goods, Mac vs. Windows, and El Farol bar games. The Mac vs. Windows game~\cite{harrington2009games} models the network effect phenomena (the value of a product increases as the number of users increases) commonly observed in economics, while the El Farol bar game~\cite{challet1998minority} models the congestion effect. We compared the expected probability of taking an action derived by our approach with the average probability of taking that action observed in the agent-based simulations. Due to space constraints, the precise game configuration details, simulation setup, and instantiation of Theorem 1 for $Q$-learning and IGA, are provided in the appendix.

\begin{figure}
\begin{subfigure}[t]{0.245\linewidth}
\centering
  \includegraphics[width=\textwidth]{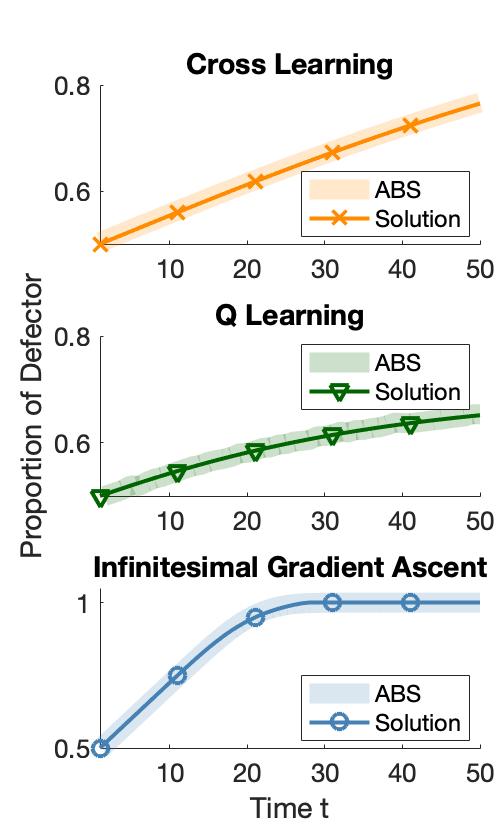}
  \small
   \vspace{-0.6cm}
  \caption{\scriptsize Public goods game.}
  \label{fig:pd}
 \end{subfigure}
\begin{subfigure}[t]{0.245\linewidth}
\centering
  \includegraphics[width=\textwidth]{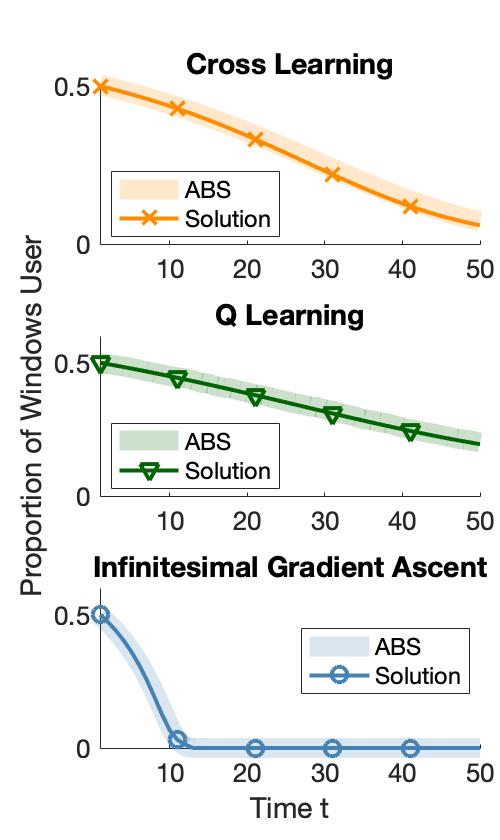}
  \small
  \vspace{-0.6cm}
  \caption{\scriptsize Mac vs. Windows game\\(critical mass for Mac initially reached).}
  \label{fig:pd}
  \end{subfigure}
\begin{subfigure}[t]{0.245\linewidth}
\centering
  \includegraphics[width=\textwidth]{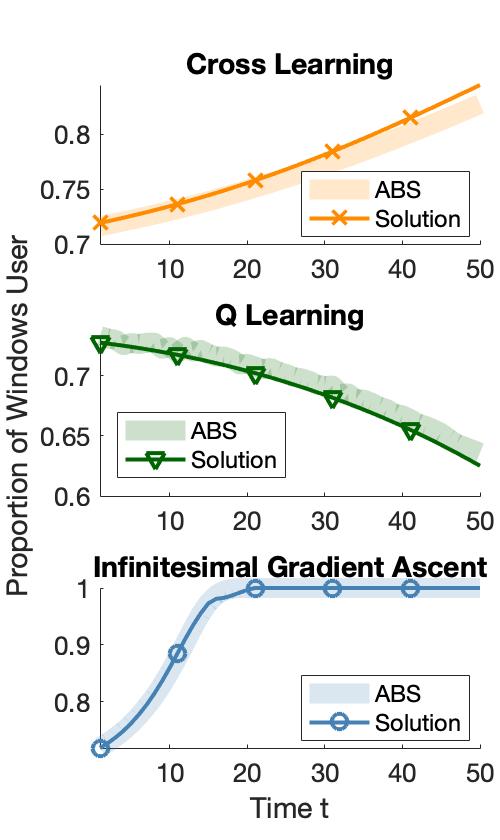}
  \small
  \vspace{-0.6cm}
  \caption{\scriptsize Mac vs. Windows game\\(critical mass for Mac initially \emph{not} reached).}
  \label{fig:pd}
    \end{subfigure}
    \begin{subfigure}[t]{0.245\linewidth}
\centering
  \includegraphics[width=\textwidth]{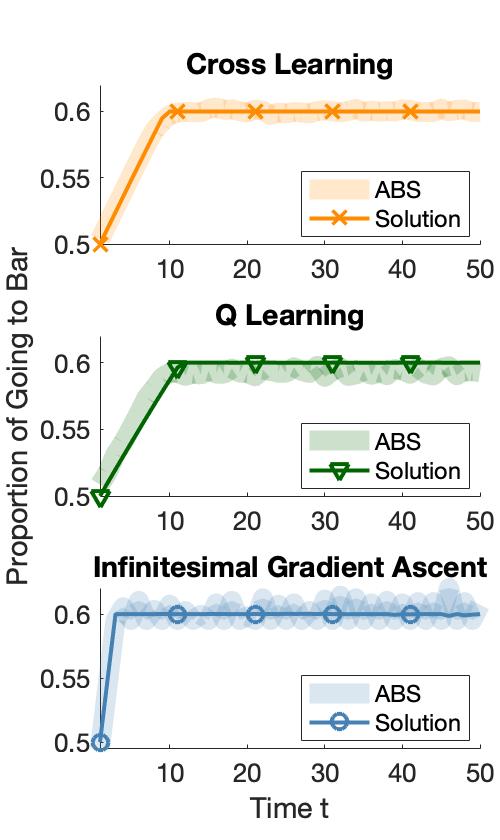}
  \small
   \vspace{-0.6cm}
  \caption{\scriptsize El Farol bar game.}
  \label{fig:pd}
    \end{subfigure}
    \\
 \begin{subfigure}[t]{\linewidth}
\centering
 \includegraphics[width=0.9\textwidth]{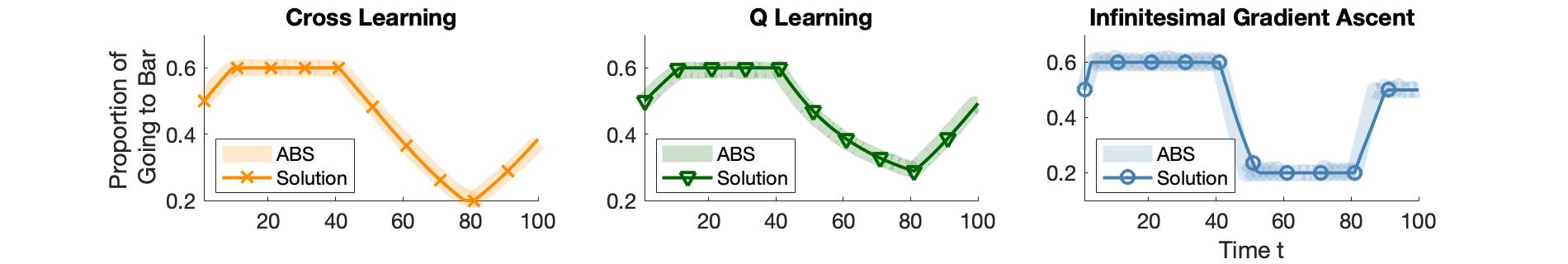}
  \small
     \vspace{-0.4cm}
  \caption{\scriptsize El Farol bar game with time-varying threshold of crowdedness. }
  \label{fig:pd}
    \end{subfigure}
  \caption{ The population dynamics derived by our approach versus that observed in agent-based simulations. The lines with markers indicate the expected probability of taking an action given by our approach. The shaded lines indicate the average probability of taking that action, which is averaged over $100$ simulations where each population  comprised $1,000$ agents.}
  \vspace{-0.5cm}
\end{figure}

\paragraph{Results and Analysis.} As Figure 2 shows, 
our approach well describes the different patterns of evolution arising in the different scenarios. These results validate Theorem 1 holds for these learning methods and population games. In the following, we report several salient findings.
More fleshed out descriptions can be found in the appendix.

In the public goods game, we observed that for a population of $Q$-learning agents, the expected probability of defection eventually stabilises around $0.7$ (Figure 2(a)). This indicates a population of $Q$-learners may not fully evolve to defect despite it being the unique Nash equilibrium state. 

In the Mac vs. Windows game,  Figure 2(b) shows that every agent will become a Mac user, if a critical mass for the prevalence of Mac is initially reached.
Symmetrically, Figure 2(c) shows that for Cross learning and IGA, every agent will become a Windows user, if the initial proportion of Mac users is slightly less than the critical mass. However, this does not hold for $Q$-learning, suggesting that a population of $Q$-learning agents is less sensitive to initial conditions.

Finally, in the El Farol bar game (Figure 2(d)), our approach shows that the expected probability of going to the bar eventually stabilises around $60\%$ (the threshold of crowdedness).
Figure 2(e) illustrates a more complex scenario where the threshold of crowdedness was decreased from $60\%$ to $20\%$ at time $t=40$, and then increased again to $50\%$ at time $t=80$.
Even under this more difficult setting, our approach well-describes the population dynamics for each learning method.

\section*{Broader Impact}
The past few years have witnessed the dramatic growth of the RL literature, and also in the sophistication of RL methods.
While it is common to treat RL as a black-box tool, there is a need to develop a better understanding of these algorithms.
Consider social robots that assist the blind, unmanned vehicles that avoid pedestrians, and autonomous agents that trade in financial markets; understanding how policies and decisions are shaped by RL methods in these scenarios is crucial. Put another way, can we \emph{trust} these methods?  
This issue will become unavoidably more salient as intelligent agents become more ubiquitous in the future.

Our work---providing a unified framework for formulating the evolutionary dynamics of independent RL methods in population games---makes one step towards addressing this issue.
Specifically, we illustrate through a case study that instantiating Theorem 1 for a particular learning method helps us identify qualitatively different patterns of evolution, perform steady state analysis, and understand the trend of expected behaviours in a population.
Theorem 1 may have broader application, such as selecting an algorithm for a specific problem, parameter tuning, and shedding light on the design of a new algorithm, which have been identified as the potential benefits of understanding RL dynamics~\cite{bloembergen2015evolutionary}.

On the flip side, the misuse of our approach may cause a negative impact.
Directly applying the PDE (Equation \ref{eq:mainresult}) to other scenarios that violate one of the assumptions stipulated in Theorem 1 may result in misleading and error prone conclusions for stakeholders.
We thus caution the use of our approach for heterogeneous or structured agent populations, and for joint-action learning methods (e.g., Nash-$Q$~\cite{hu2003nash}, minimax-$Q$~\cite{littman1994markov} and Friend-or-Foe $Q$~\cite{littman2001friend}).
A promising direction for future research is to study  evolutionary learning dynamics when the aforementioned scenarios are considered.

\section*{Appendix A. Details about the Derivation}
In the following, more details about the derivation are presented.
In Appendix A.1, we show how we reach the main result (Equation 5 in Theorem 1) from Equation 12 of the main text.
In Appendix A.2, we provide detailed explanations about how we instantiate the probabilities 
$\mathscr{T}(\mathbf{x}, \Delta \mathbf{x},t)$ and $\mathscr{T}(\mathbf{x}, -\Delta \mathbf{x},t)$, which are used in the master equation.
In Appendix A.3, we discuss the implication of Equation 8 of the main text in detail.

\subsection*{A.1 Derivation from Equation 12 to Equation 5}
In the main text, we mention that Equation 5 in Theorem 1 can be derived from Equation 12. In the following, we show the intermediate steps.
\begin{align}
\frac{\partial p(\mathbf{x},t)}{\partial t} & =
\int  \Delta{\mathbf{x}} \cdot \nabla_{\mathbf{x}}[\delta(f_{\theta}(\mathbf{x}, g(\mathbf{x},t) ,\mathcal{I}(t),t) + \Delta \mathbf{x}) p(\mathbf{x},t)]  d\Delta \mathbf{x}. \\
&=  \int \sum_{j=1}^{d} \Delta x_j \frac{\partial [ \delta(f_{\theta}(\mathbf{x}(t),g(\mathbf{x},t)  ,\mathcal{I}(t),t) + \Delta \mathbf{x})  p(\mathbf{x},t)] } {\partial x_j} d\Delta \mathbf{x} \\
&=  \sum_{j=1}^{d} \frac{\partial}{\partial x_j}\int \Delta x_j   \delta(f_{\theta}(\mathbf{x}(t),g(\mathbf{x},t) ,\mathcal{I}(t),t) + \Delta \mathbf{x})  p(\mathbf{x},t) d\Delta \mathbf{x} \\ \label{eq:1}
&=  \sum_{j=1}^{d} \frac{\partial}{\partial x_j} [ p(\mathbf{x},t) \int \Delta x_j   \delta(f_{\theta}(\mathbf{x}(t),g(\mathbf{x},t) ,\mathcal{I}(t),t) + \Delta \mathbf{x}) d\Delta \mathbf{x}] \\ \label{eq:2}
&= - \sum_{j=1}^{d} \frac{\partial} {\partial {x_j}} [p(\mathbf{x},t) f_{\theta,j}(\mathbf{x}(t),g(\mathbf{x},t) ,\mathcal{I}(t),t)].
\end{align}
From the above Equation \ref{eq:1} to \ref{eq:2}, 
when $\Delta \mathbf{x} =- f_{\theta}(\mathbf{x}(t),r(t),\mathcal{I}(t),t)$ in which $\Delta x_j = -f_{\theta,j}(\mathbf{x}(t),r(t),\mathcal{I}(t),t)$, the delta function 
$\delta(f_{\theta}(\mathbf{x}(t),r(t),\mathcal{I}(t),t) + \Delta \mathbf{x}) \to +\infty$.
Otherwise, the delta function equals $0$.
Therefore, the integral in Equation \ref{eq:1} equals $-f_{\theta,j}(\mathbf{x}(t),r(t),\mathcal{I}(t),t)$.

\subsection*{A.2 Instantiation of Probabilities $\mathscr{T}(\mathbf{x}, \Delta \mathbf{x},t)$ and $\mathscr{T}(\mathbf{x}, -\Delta \mathbf{x},t)$}
For a single agent $i$, at given time $t$, its change in its critical parameters is given by the function $f_\theta$ (Equation 3 in Theorem 1).
Because every agent in the population adopts the same function $f_\theta$,
at given time $t$, the change in a particular vector $\mathbf{x}$ of critical parameters does \emph{not} vary for different agents, and is uniquely given by $f_{\theta}(\mathbf{x},g(\mathbf{x},t),\mathcal{I}(t),t)$.
Thus, we can characterise the probability $\mathscr{T}(\mathbf{x}, \Delta \mathbf{x},t)$ of a transition from $\mathbf{x}$ to $\mathbf{x}+\Delta\mathbf{x}$ by a delta function 
\begin{equation}
\begin{aligned}
\mathscr{T}(\mathbf{x}, \Delta \mathbf{x},t)  \triangleq \delta(f_{\theta}(\mathbf{x},g(\mathbf{x},t),\mathcal{I}(t),t) - \Delta \mathbf{x}).
\end{aligned}
\end{equation}
The probability  $\mathscr{T}(\mathbf{x}, \Delta \mathbf{x},t)$ is non-zero if and only if the change in the vector $\mathbf{x}$ of critical parameters, given by $f_{\theta}(\mathbf{x},g(\mathbf{x},t),\mathcal{I}(t),t)$, is exactly $\Delta \mathbf{x}$.
Similarly, the probability $\mathscr{T}(\mathbf{x}, -\Delta \mathbf{x},t)$ of a transition from $\mathbf{x}$ to $\mathbf{x}-\Delta\mathbf{x}$ can be defined by a 
delta function 
\begin{equation}
\begin{aligned}
\mathscr{T}(\mathbf{x}, -\Delta \mathbf{x},t)  \triangleq \delta(f_{\theta}(\mathbf{x},g(\mathbf{x},t),\mathcal{I}(t),t) + \Delta \mathbf{x}).
\end{aligned}
\end{equation}
The probability  $\mathscr{T}(\mathbf{x}, -\Delta \mathbf{x},t)$ is non-zero if and only if the change in the vector $\mathbf{x}$ of critical parameters, given by $f_{\theta}(\mathbf{x},g(\mathbf{x},t),\mathcal{I}(t),t)$, is exactly $-\Delta \mathbf{x}$.

Note that the integral of a delta function is $1$. Hence, we have
\begin{equation}
\begin{aligned}
\int \mathscr{T}(\mathbf{x}, \Delta \mathbf{x},t)  d \Delta \mathbf{x}= \int \delta(f_{\theta}(\mathbf{x},g(\mathbf{x},t),\mathcal{I}(t),t) - \Delta \mathbf{x})  d \Delta \mathbf{x}=1.
\end{aligned}
\end{equation}
and 
\begin{equation}
\begin{aligned}
\int \mathscr{T}(\mathbf{x}, -\Delta \mathbf{x},t)  d \Delta \mathbf{x}= \int \delta(f_{\theta}(\mathbf{x},g(\mathbf{x},t),\mathcal{I}(t),t) + \Delta \mathbf{x})  d \Delta \mathbf{x}=1.
\end{aligned}
\end{equation}

\subsection*{A.3 Homogeneity required by Equation 8}
In the main text, we show that to ensure the time evolution of the $p(\mathbf{x},t)$ is asymptotically that of the empirical distribution, the following equation should hold:
\begin{equation}
T(\mathbf{x}|\mathbf{x}',t) = \mathcal{T}^{(i)}(\mathbf{x}|\mathbf{x}',t), \qquad \forall i\in \mathcal{N}.
\end{equation}
This equation requires that at given time $t$, the probability of a transition for any agent should be uniquely determined by the critical parameters from which and to which the agent transits, and be independent of their identity.
In other words, agents should be homogeneous, in that they are distinguished only by their critical parameters.
Note that for a single agent $i$, its change in the critical parameters is given by the parameterised function $f_\theta$ (Equation 3 in Theorem 1).
Hence, as stipulated in Theorem 1, there should exist the same function $f_\theta$ for every agent the population, in which the non-critical parameters $\theta$ should be also the same.
Moreover, the function $f_\theta$ depends on the critical parameters, and also on the immediate reward and the additional information set.
To ensure the homogeneity, there should exist a function (function $g$ in Theorem 1) that maps a vector of critical parameters to an immediate reward, such that at given time $t$, agents having the same critical parameters receive the same immediate reward.
In addition, agents having the same critical parameters should make use of the same additional information.

\section*{Appendix B. Instantiations of Theorem 1}
In the following, we briefly describe $Q$-learning with Boltzmann exploration and infinitesimal gradient ascent, and show how Theorem 1 can be instantiated with these two learning methods.
\subsection*{B.1 $Q$-learning with Boltzmann exploration}
\paragraph{Description of the method.}
In a single-state (or stateless) MDP with a set $\mathcal{A}$ of $k$ available actions, a $Q$-learning agent \cite{watkins1992q} maintains a vector of $Q$-values $\mathbf{Q}(t)=[Q_1(t),\ldots ,Q_k(t)]^\top$, each of which $Q_i(t)$ estimates the expected reward of using an action $a_i \in \mathcal{A}$ at time $t$.
Suppose that at time $t$, an agent takes action $a_j$ and receives an immediate reward $r(t)$ accordingly.
This agent will update the $Q$-value for each action $a_i$ 
as follows: 
\begin{equation}
Q_{i}(t+1)  = 
\begin{cases}
(1-\alpha)  Q_{i}(t)  +\alpha  r(t) & \text{if } a_i=a_j \\
Q_{i}(t) & \text{else}
\end{cases},
\qquad \forall a_i\in \mathcal{A},
\label{eq:newq}
\end{equation}
where $\alpha$ is the learning rate. 
Note that the term estimating the optimal reward after state transition is dropped, since agents stay in the same state \cite{hu2019modelling,kianercy2012dynamics,Gomes2009dynamic,wunder2010classes}.

The agent then updates its policy based on the updated $Q$-values. 
Let $\boldsymbol \pi(t+1)$ be its mixed-strategy policy at time $t+1$, such that $\boldsymbol \pi(t+1)=[\pi_1(t+1), \ldots, \pi_k(t+1)]^\top$ in which each element $\pi_i(t+1)$ is the probability of taking action $a_i$ at time $t+1$.
A $Q$-learning agent using Boltzmann exploration will update the policy as follows:
\begin{equation}
\pi_{i}(t+1)  = \frac{e^{\tau Q_i(t+1)}} {\sum_{j: a_j \in \mathcal{A}}{e^{\tau Q_{j}(t+1)}}}, \qquad \forall a_i\in \mathcal{A},
\end{equation}
where $\tau$ is the Boltzmann exploration temperature. 
A larger value of $\tau$ indicates the fewer exploration for individual agents.
When $\tau \to \infty$, agents take the action with the highest $Q$-value in probability $1$.

\paragraph{Instantiation.}
We consider an agent's $Q$-values $\mathbf{Q}(t) \in \mathbb{R}^k$ to be its critical parameters.
It is shown \cite{hu2019modelling} that the differential equation that describes the time evolution of an agent's $Q$-values in stateless MDP is given by
\begin{equation}
\frac{dQ_{i}(t)}{dt}  = \alpha \frac{e^{\tau Q_i(t)}} {\sum_{ j: a_j \in \mathcal{A}}{e^{\tau Q_{j}(t)}}}  [r(t)- Q_{i}(t) ], \qquad \forall a_i\in \mathcal{A}.
\label{eq:q}
\end{equation}
By Theorem 1, we have the following partial differential equation:
\begin{equation}
\begin{aligned}
\frac{\partial p(\mathbf{Q},t)}{\partial t}  &= -\sum_{i:a_i\in \mathcal{A}}\frac{\partial }{\partial Q_i} \left [ p(\mathbf{Q},t) \frac{dQ_{i}(t)}{dt} \right ], \\
\end{aligned}
\label{eq:qfpk}
\end{equation}
where $p(\mathbf{Q},t)$ is intuitively interpreted as the proportion of agents having the particular vector $\mathbf{Q}=[Q_1,\ldots, Q_k]^\top$ of $Q$-values in the population at time $t$.

\subsection*{B.2 Infinitesimal Gradient Ascent}
\paragraph{Description of the method.}
Infinitesimal Gradient Ascent \cite{singh2000nash} (IGA) was originally designed for two-player-two-action games.
Consider a set $\mathcal{A}$ of $2$ available actions.
An IGA agent maintains a vector of policy $\boldsymbol{\pi}(t)= [\pi_1(t),\pi_2(t)]^\top$, in which each element $\pi_i(t)$ is the probability of using an action $a_i \in \mathcal{A}$. 
At time $t$, if the agent takes action $a_j$ and receives an immediate reward $r(t)$, then it will update its policy $\boldsymbol{\pi}(t)$, as follows:
\begin{equation} 
\pi_i(t+1) = \pi_i(t) + \alpha \frac{\partial V(\boldsymbol {\pi},t)}{\partial \pi_i}, \qquad \forall a_i\in \mathcal{A},
\label{eq:iga}
\end{equation}
where $\alpha$ is the step size (or learning rate), and $V(\boldsymbol {\pi},t)$ denotes the value function that maps a policy $\boldsymbol {\pi}$ to its expected reward at time $t$.
IGA assumes agents have access to the value function.
We extend this method by assuming that the reward of taking each action is known to agents, such that the value function $V(\boldsymbol {\pi},t)$ is defined as
\begin{equation} 
V(\boldsymbol {\pi},t) = \sum_{i:a_i\in \mathcal{A}} \pi_i \times r_i(t), 
\label{eq:iga}
\end{equation}
where $r_i(t)$ is the reward of taking action $a_i$ at time $t$.
Therefore, an IGA agent will update its policy by taking steps in the direction of the gradient of its expected reward.

\paragraph{Instantiation.}
For two-action games, we consider an IGA agent's critical parameter to be its probability $\pi_1(t)$ of taking action $a_1$.
It is shown \cite{kaisers2012common} that the differential equation that describes the change in the probability of taking an action (which is action $a_1$ here) is given by
\begin{equation}
\begin{aligned}
\frac{d\pi_1(t)}{dt}  &= \alpha \frac{\partial V(\boldsymbol {\pi},t)}{\partial \pi_i}. 
\end{aligned}
\end{equation}
By Theorem 1, we have the following partial differential equation:
\begin{equation}
\begin{aligned}
\frac{\partial p(\pi_1,t)}{\partial t}  &= -\frac{\partial }{\partial \pi_1} \left [ p(\pi_1,t) \frac{d\pi_1(t)}{dt} \right ], \\
\end{aligned}
\label{eq:qfpk}
\end{equation}
where $p(\pi_1,t)$ is intuitively interpreted as the proportion of agents having the particular probability $\pi_1$ of taking action $a_1$ in the population at time $t$.

\section*{Appendix C. Details about the Experiments}
In the following, we present the experimental settings, precise game configurations, and the detailed results across different learning methods and population games.
\subsection*{C.1 Experimental Settings}
For validation, we compare the expected probability of taking an action derived by our approach with the average probability of taking that action observed in agent-based simulations.
For the learning rate (or step size) $\alpha$ in $Q$-learning and IGA, we set it to be $0.05$.
For Boltzmann exploration temperature $\tau$ in $Q$-learning, we set it to be $2$.
We note that Cross learning requires the rewards to range in $[0,1]$.
To get rid of this requirement, we impose a small update step with the size $\alpha=0.01$ (by multiplying the update term by $\alpha=0.01$), following the common practice \cite{bloembergen2015evolutionary}.

Unless stated otherwise, for Cross learning and IGA, we assume the initial probability of taking the first action for each agent is distributed according to the truncated normal distribution $\mathsf{N}(0.5, 0.1^2)$.
For $Q$-learning, we assume the initial $Q$-values for the two actions are both distributed according to a truncated normal distribution $\mathsf{N}(({r_{\max}+r_{\min}})/{2}, 0.4^2)$, in which $r_{\max}$ and $r_{\min}$ denote the maximal and the minimal rewards of a population game, respectively.
As a result, agent populations use different learning methods though, initially, the expected probability (or the average probability in simulations) of taking the first action is all around $0.5$.
In other words, agents generally choose their actions randomly in the beginning.

In the agent-based simulations, we consider a population of $1,000$ agents.
To smooth out randomness, we run $100$ simulations for each setting.

\subsection*{C.2 Game Configurations and Detailed Results}

\paragraph{Public Goods Game.} 
As we mentioned in Section 4 of the main text, in this game, each agent has two actions: defect (denoted by $a_1$) or  cooperate (denoted by $a_2)$. 
An agent will receive a reward $r_1(t)=1.5\psi$ if it defects , and a reward $r_2(t)=1.5\psi-0.5$ if it cooperates, in which $\psi$ is the current proportion of cooperating agents.
Because defection always yields a higher reward than cooperation, the unique Nash equilibrium in this game is that every agents converge to defect.

Not surprisingly, our approach shows in Figure 2(a) of the main text that all of the agent populations develop a tendency to take defection.
However, our approach also indicates that to what degree agents develop such tendency vary greatly for the use of different methods.
In fact, for a population of $Q$-learning agents, the expected probability of taking defection eventually stabilises around $0.7$.
This suggests that a population of $Q$-learning agents may not evolve into the full convergence to defect despite it being the unique Nash equilibrium state.

\paragraph{Mac vs. Windows Game.}
This game \cite{harrington2009games} models the network effect phenomena commonly observed in economics.
When a network effect is present, the value of a product or service increases as the number of users increases.
For example, as there are more Windows users, people can share files with a larger community, and companies are more attracted to developing software for Windows users. This will in turn create a positive reinforcement, resulting in an even larger userbase.

In this game, there are two actions, namely, use Mac (denoted by $a_1$) and use Windows (denoted by $a_2$). 
Let $\psi$ be the proportion of Mac users at time $t$.
The immediate reward function is given as: $r_1(t)=0.5+1.5 \psi$ and $r_2(t)=1.5(1-\psi)$.
By this reward function, Mac is inherently more superior, however, using Windows will receive a higher reward if less than $\frac{1}{3}$ of agents in the population use Mac. 
In other words, $\frac{1}{3}$ of Mac users is the critical mass for the prevalence of Mac.
As shown in Figure 2(b) of the main text, our approach shows that every agent will become a Mac user, if initially around half of the agents in the population are Mac users.

We then consider another initial situation, such that there are around $28\%$ of Mac users (slightly less than the critical mass) in an agent population for each learning method. 
Specifically,
for Cross learning and IGA, we assume the initial probability of taking the first action (using Mac) for each agent is distributed according to the truncated normal distribution $ \mathsf{N}(0.28, 0.1^2)$.
For $Q$-learning, we assume the initial $Q$-values for the first and the second actions are distributed according to the truncated normal distributions $\mathsf{N}(0, 0.1^2)$ and $ \mathsf{N}(0.5, 0.1^2)$, respectively.
It is shown in Figure 2(c) that for cross learning and IGA, every agent becomes a Windows user.
However, for $Q$-learning, every agent becomes a Mac user.
This suggests that compared with cross learning and IGA, a population of $Q$-learning agents is less sensitive to the initial conditions, which is critical in population games with a network effect.

\paragraph{El Farol Bar game.} 
This game \cite{arthur1994complexity} is an extensively-studied model \cite{whitehead2008farol} for the congestion effect phenomena widely exist in economics and daily life.
In this game, there are two actions: go to the El Farol Bar (denoted by $a_1$), and stay home (denoted by $a_2$). 
Agents staying home always receive no reward, i.e., $r_2(t)=0$. For agents going to the bar, an agent will receive a reward $r_1(t)=1$ if currently less than $60\%$ of the agents in a population go to the bar. However, it will receive a punishment $r_1(t)=-1$ otherwise, because the bar is too crowded.
As shown in Figure 2(d) of the main text, our approach shows that the expected probability of going to bar eventually stabalises around $60\%$.

To further validate our approach, we deliberately make the game more complex, so that the population dynamics become more unpredictable. 
We decrease the threshold of crowdedness from $60\%$ to $20\%$ at time $t=40$, and then increase it again to $50\%$ at time $t=80$.
It is shown in Figure 2(e) that under this more difficult setting, our approach still well describes the population dynamics for each learning method.
Moreover, our approach also shows that IGA agents change their probability of going to bar more timely due to the change in the threshold of crowdedness.
This suggests that IGA is more responsive than $Q$-learning and cross learning in population games with a congestion effect.

\bibliographystyle{plain}
\bibliography{ref}

\end{document}